\documentclass[prl, twocolumn, amsmath,floatfix,unsortedaddress,superscriptaddress]{revtex4-1}

\usepackage{graphicx}
\usepackage{natbib}

\newcommand{\sciexp}[2]{{#1}\ensuremath{\,\times\,10^{#2}}}
\newcommand{\pow}[1]{\ensuremath{^{{#1}}}}

\newcommand{\UNH}{Space Science Center, University of New Hampshire, Durham, NH, 03824}
\newcommand{\LLE}{Laboratory for Laser Energetics, University of Rochester, Rochester, NY 14623}
\newcommand{\FSC}{Fusion Science Center for Extreme States of Matter, University of Rochester, Rochester, NY 14623}
\newcommand{\PPPL}{Department of Astrophysical Sciences and Princeton Plasma Physics Laboratory, Princeton, NJ 08543}

\begin{document}

\begin{abstract}

Filamentation due to the growth of a Weibel-type instability was observed in the
interaction of a pair of counter-streaming, ablatively-driven plasma flows,
in a supersonic, collisionless regime relevant to astrophysical collisionless shocks.
The flows were created by irradiating a pair of opposing plastic (CH) foils 
with 1.8~kJ, 2-ns laser pulses  on the \textsc{omega ep} laser system.  
Ultrafast laser-driven proton radiography was used to image the Weibel-generated
electromagnetic fields.  The experimental observations are in
good agreement with the analytical theory of the Weibel instability and
with particle-in-cell simulations.

\end{abstract}

\title{Filamentation instability of counter-streaming laser-driven plasmas}
\author{W. Fox}
\affiliation{\UNH}
\author{G. Fiksel}
\affiliation{\LLE}
\affiliation{\FSC}
\author{A. Bhattacharjee}
\affiliation{\PPPL}
\author{P.-Y. Chang}
\affiliation{\LLE}
\affiliation{\FSC}
\author{K. Germaschewski}
\affiliation{\UNH}
\author{S. X. Hu}
\affiliation{\LLE}
\author{P. M. Nilson}
\affiliation{\LLE}
\affiliation{\FSC}
\date{\today}

\maketitle

Astrophysical shock waves play diverse roles, including
energizing cosmic rays in the blast waves of astrophysical explosions 
\cite{AckermannScience2013, *BellMNRAS1978, *BlandfordApJ1978},
and generating primordial magnetic fields during the 
formation of galaxies and clusters
\cite{RyuScience2008, *KulsrudApJ1997}.
These shocks are typically collisionless, and require collective
electromagnetic fields \cite{Sagdeev1966}, as 
Coulomb collisions alone are too weak to sustain shocks 
in high-temperature astrophysical plasmas.  
The class of Weibel-type instabilities 
\cite{WeibelPRL1959,FriedPoF1959,DavidsonPoF1972}
(including the classical Weibel and current-filamentation instabilities)
is one such collective mechanism that has been
proposed to generate a turbulent magnetic field in the shock front and thereby mediate shock formation
in cosmological shocks \cite{SchlickeiserApJ2003} and blast wave shocks in 
gamma ray bursts \cite{MedvedevApJ1999, KatoApJ2007, SpitkovskyApJL2008}
and supernova remnants \cite{KatoApJL2008}.
These instabilities generate magnetic field \textit{de novo} by tapping 
into non-equilibrium features in the electron and ion distributions functions.
The classical form of the Weibel instability is driven by temperature anisotropy \cite{WeibelPRL1959}, but counterstreaming
ion beams, as occurs in the present context, provides an equivalent drive mechanism \cite{DavidsonPoF1972}.
A related current filamentation instability of relativistic electron beams \cite{CalifanoPRE1998,*PukhovProgPhys2003}
has also previously been observed in experiments driven by ultraintense lasers \cite{JungPRL2005, *MondalPNAS2012}.

We report experimental identification an ion-driven Weibel-type instability generated in the interaction of two
counterstreaming laser-produced plasma plumes.
A pair of opposing CH targets was irradiated by kJ-class laser pulses 
on the OMEGA EP laser laser system, driving a pair of ablative flows toward the collision
region at the midplane between the two foils.  Due to the 
long mean-free-path between ions in opposing streams,
the streams interpenetrate, establishing supersonic counterstreaming conditions in
the ion populations, while the electrons form a single thermalized cloud.
Meanwhile, the plasma density is also sufficient so
that the the ion skin depth $d_i = (m_i / \mu_0 n e^2)^{1/2}$,
is much smaller than the system size $L$.  These conditions allow
the growth of an ion-driven Weibel instability, for which $d_i$ is the characteristic wavelength
 \cite{TakabePPCF2008, ParkHEDP2011, DrakeApJ2012}.
 The Weibel-generated electromagnetic fields were observed with an ultrafast
proton radiography technique \cite{BorghesiPRL2004}, and identified through good
agreement with analytic theory \cite{DavidsonPoF1972} and particle-in-cell simulations, discussed below.

\begin{figure}
\includegraphics[width=3.375in]{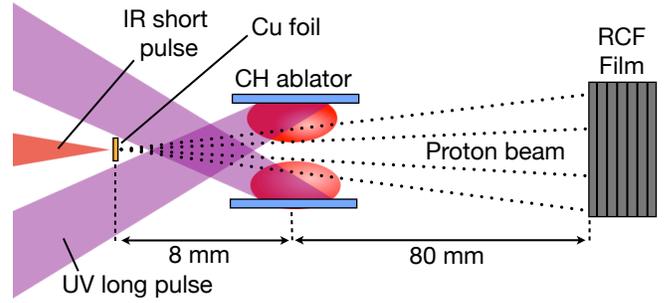}
\caption{Experimental setup: Plasma plumes are ablated from
a pair of CH targets, separated by $2L$ = 4.5~mm, which collide and interact in the midplane.
The electromagnetic fields formed due to instabilities in the interaction
were radiographed with a laser-driven proton beam and imaged
onto radiochromic film (RCF).}
\label{FigSetup}
\end{figure}

\begin{figure*}[t]
\includegraphics[width=170mm]{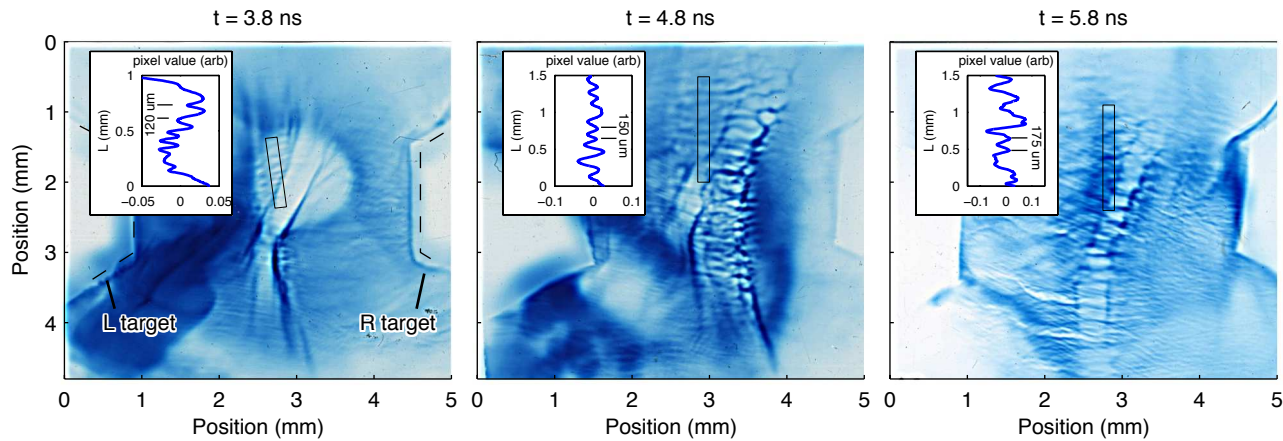}
\caption{Radiochromic film images of the development of a striated
instability at the midplane at 3.8~ns, 4.8~ns, and 5.8~ns relative to the 
start of the main driver laser pulse.  The film records fluence of protons with energy of
order 5--10~MeV, with darker features reflecting greater proton
fluence.  
The striated features are well-reproduced on neighboring film in the stack,
reflecting the strong-focusing of the diagnostic proton beam \cite{KuglandRSI2012}.
The insets show 1-d traces of proton intensity along the long-axis of the regions
denoted in the film, averaged over the short direction, from which 
a quadratic background proton variation was subtracted
to focus on the fluctuations.  Typical filament wavelengths of 
100--150~$\mu$m at 3.8 and 4.8~ns expand to wavelengths near 250~$\mu$m
at 5.8~ns.  The distances on the axes are those
in the nominal object plane, which is perpendicular to the axis of the proton beam 
and the targets, and which goes through the center of the UV drive laser foci. 
The image has been sharpened in post-processing to emphasize the striations.
(Raw images available in supporting material).
}
\label{FigWeibelFilm}
\end{figure*}

Figure~\ref{FigSetup} shows a schematic of the experiments.
A pair of opposing plastic (CH) targets, separated by 2$L$ = 4.5~mm, were 
each irradiated with  1.8~kJ, 2~ns laser pulses at a wavelength of 0.351~$\mu$m.
The laser pulses irradiated the targets at highly oblique angle of incidence ($\theta \approx 74^\circ$),
leading to highly-elliptical focal spots ($e \approx 3.5$).  The beam foci had a minor diameter of 900~$\mu$m,
and used distributed phase-plate (DPP) beam smoothing, for on-target laser intensities near \sciexp{5}{13}~W/cm$^2$.
The laser setup was similar to the recent interacting plume experiments of Kugland \textit{et al} \cite{KuglandNPhys2012}, which
observed large-scale ``self-organized'' plasma structures, except for
smaller separation of the targets and the use of broader laser foci and DPP phase plate beam smoothing,
which may limit the density and decrease the magnitude of self-generated magnetic fields.

The electromagnetic fields formed in the interaction region were probed using 
an ultrafast diagnostic proton beam \cite{BorghesiPRL2004}, generated with a third,
high-intensity laser pulse (1.053~$\mu$m, 800J, 10 ps) focused to $>10^{18}$~W/cm$^2$, 
irradiating a thin Cu disk 8~mm from the interaction region with a focal spot of about 25~$\mu$m.
This created a uniform and laminar point source of protons,
with a distribution of energies of order 10~MeV via the target-normal sheath acceleration
mechanism \cite{WilksPoP2001}.
 A Ta shield prevented the ablated plasmas from
interfering with the proton beam formation.
The protons were detected with a stack of radiochromic film 
80~mm from the interaction region, for a geometrical magnification of 11, with proton energies resolved
in the film stack by their varying Bragg peaks.
Figure~\ref{FigWeibelFilm} shows the development of the instability in
a sequence of radiographic images, taken over
multiple experimental shots by
varying the timing of the proton beam with respect to the main driver lasers.
The two ablator targets are visible at the left and right of the images.
The sequence of proton images reflects the electromagnetic structures formed in the plasma
as the two plasma plumes interpenetrate. 

The first image, at $t =$~3.8~ns relative to the start of the driver pulse, shows a
prominent and sharp``X''-like structure at the midplane, with the protons deflected
into pairs of thin lines,
reminiscent of the caustic proton structures
observed in experiments in a similar laser-energy regime with larger initial target separation \cite{KuglandRSI2012, KuglandNPhys2012}.
However, for the present discussion, we focus on the
filamentary instability visible above the ``X'' structure.
In the inset, a 1-d cut 
shows a quasi-periodic variation in proton fluence, which is 
still relatively weak at this time, suggesting that this frame
catches the linear growth of the instability just as it becomes measurable
with the proton diagnostic.
The instability has a wavelength of about 100--120~$\mu$m transverse
to the counterstreaming flows, and
only a single wavelength or eigenmode-like structure parallel to the flows, 
over a width of about 500~$\mu$m.
The second image at $t$~=~4.8~ns shows substantial growth of the instability,
with much larger variation in proton fluence, and with wavelengths transverse to the counterstreaming flows
of order 120--150~$\mu$m.  The proton fluctuations still retain a filamentary
character with longer correlation lengths parallel to the flows; 
however, the growth of multiple modes has replaced the clean single-eigenmode structure from 3.8~ns.
The third image, at $t = $~5.8~ns, shows the non-linear evolution of the 
instability.  There remains a primary ``spine'' of instability along the midplane.  However, 
the typical wavelengths are even longer in the vertical direction, and the
ratio of wavelengths parallel and perpendicular to the flows is closer to unity.
This can reflect the non-linear evolution of the instability, including
coalescence of filaments or the dilation of filaments
frozen into the outflows of plasma from the stagnation point at the midplane.

Radiation-hydrodynamics simulations with the 
DRACO code \cite{RadhaPoP2005, HuPoP2010} provide baseline predictions for interpreting the observations.
The simulations show quasi-isothermal plasma ablation \cite{MannheimerPoF1982}
with initial electron temperatures $T_e$ near 800~eV driving
a supersonic ablation stream flowing from each target with $V \approx C_s + x/t$, 
and density $n \approx n_{ab} \exp (-\alpha x/C_st)$.
where $C_s \approx $~\sciexp{2}{5}~m/s is the sound speed, $x$ measures distance from the 
ablation surface, and $t$ time.  The ablation density $n_{ab}$ of \sciexp{7}{26}~m$^{-3}$ is reduced
compared to the critical density owing to the highly oblique laser incidence, and $\alpha$ is a factor of order 1.
The conditions at the midplane during instability growth are estimated by superposition
of single-plume DRACO simulations; for example, 
at $t=3.8~$ns, the electron density (summed over both plumes) is \sciexp{2}{25}~m$^{-3}$,
and the electron temperature is 250~eV.  The ion distribution function
consists of two counterstreaming beam components with flow
speeds $\pm V$ near \sciexp{8}{5}~m/s, with single-stream ion temperatures near 150~eV. 
Under such conditions the ion-ion inter-plume interaction is quasi-collisionless, 
with the C$^{6+}$-C$^{6+}$ mean-free-path between
the opposing streams of order 10~cm, and the C$^{6+}$-electron mean-free-path at least 4~mm 
(and likely longer if the electrons are heated in the interaction region \cite{RossPoP2012}).
Meanwhile, the electron collision frequency is faster than the dynamics we consider, 
therefore the electrons form a single thermalized population.

\begin{figure}
\includegraphics{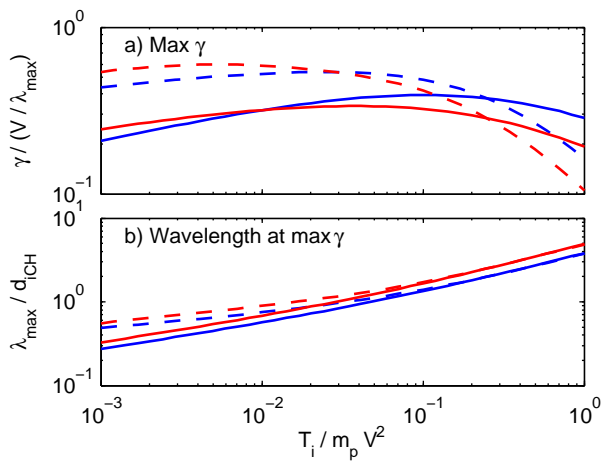}
\caption{a) Growth-rate of Weibel instability and (b) Wavelength of fastest growing mode
vs. ion temperature.  Specifically, we solve the ion-pinch dispersion relation of Davidson \textit{et al} \cite{DavidsonPoF1972}, 
using zero electron anisotropy due to the electron collisionality regime of the experiments.
The growth rate is normalized to the counter-streaming speed $V$ and wavelength of max growth for each ion temperature.
The blue curves are for a 60-40 H-C mixture close to the composition of the targets,
and red is for a 90-10 mixture reflecting fractionation.  Solid curves are for 
$T_e = T_i$, while dashed is for heated electrons with $T_e = 0.25 m_pV^2$.
To compare wavelengths on an even footing, for both compositions the wavelengths are 
normalized a nominal $d_{iCH}$ calculated with the 60-40 H-C mixture: $d_{iCH} \approx 1.29 (m_p / \mu_0 n_e e^2)^{1/2}$.}
\label{FigWeibelTscan}
\end{figure}

These collisionless, counterstreaming conditions at the midplane are requisite conditions to drive the Weibel instability,
and the observations bear many expected qualitative features of 
this instability, including localization to the overlap region and formation
of elongated filaments parallel to the ablation flows.
We have obtained quantitative agreement with the local
electromagnetic dispersion relation for the ion-driven Weibel instability \cite{DavidsonPoF1972},
which includes counterstreaming ions but a single collisionally thermalized electron population.
The dispersion relation is solved directly for the 
maximum growth rate $\gamma_{max}$ and fastest-growing wavelength $\lambda_{max}$ versus
the electron and ion temperature (shown in Fig.~\ref{FigWeibelTscan}),
which shows robust instability growth over a wide range of temperatures and possible fractionation
effects in the colliding plasmas.
Furthermore, when the growth rate is normalized to the counter-streaming speed 
$V$ and the fastest-growing wavelength for each ion temperature, 
as in Fig.~\ref{FigWeibelTscan}(a), 
the results are remarkably constant over orders of magnitude variation 
in ion temperature, demonstrating that $V/\lambda$ is the dominant scaling.
Experimentally, this is important, as recent experiments 
have observed complex heating dynamics of both
electrons and ions in similar counterstreaming conditions \cite{RossPoP2012, RossPRL2013}.
Instead, this dominant scaling enables straightforward comparison of observations and
the linear theory: the wavelengths
are measured in the radiography, and the interaction speed $V \approx C_s + L/t$ is well-constrained
due to the simple nature of the ablative flow.
For example, for $t~=~3.8~$ns, $V$ is estimated as \sciexp{8}{5}~m/s,
and DRACO predicts a single-stream ion-temperature of 150~eV.  Accounting
for possible ion heating leads us to consider a range of $T_i/m_p V^2$ from 0.01--0.1.
In this $T_i$ range, and considering the heated electron curve (dashed curves),
$\gamma \approx 0.5 V/\lambda$ .  
Assuming that the observed
wavelengths (near 150~$\mu$m) are close to the fastest growing modes,
the growth rate is then estimated at \sciexp{2--3}{9}~s\pow{-1},
which is consistent with the rapid appearance 
of the filaments on the 1~ns period separating
adjacent frames.

\begin{figure*}
\includegraphics[width=6.5in]{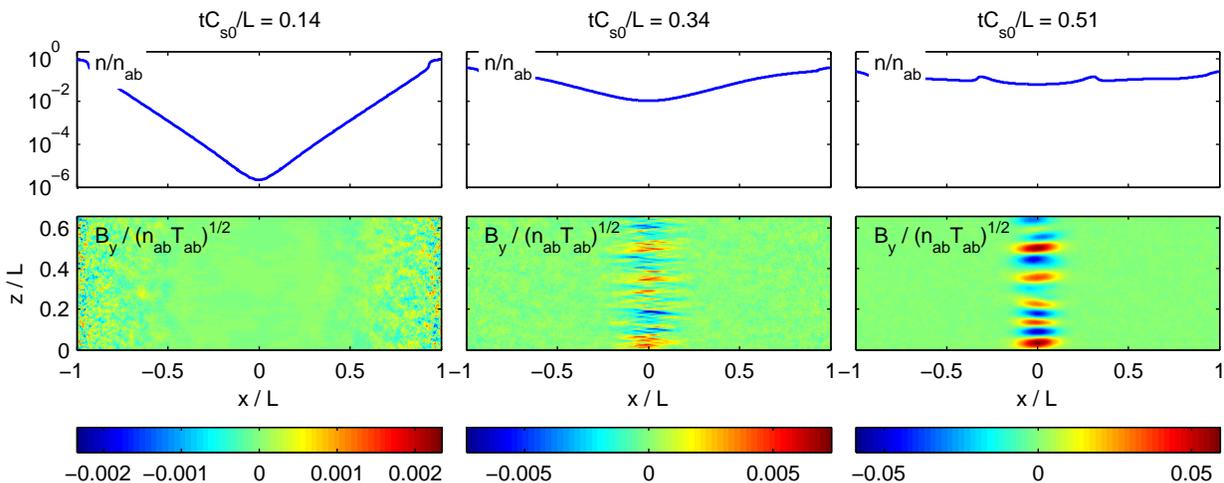}
\caption{PIC simulation of growth of Weibel filaments between counterstreaming ablation flows.  Top: evolution
of the plasma density.  Bottom: development of transverse magnetic filaments from the Weibel instability.
To generate the counterstreaming ablation-flow geometry, plasma is added dynamically 
to small volumes at the left and right boundaries for time $t=[0,t_{laser}]$.
This sets up a pair of flows with ablation-like profiles for density ($n \approx n_{ab} \exp(-\Delta x/C_s t)$) and 
velocity ($V \approx C_s + \Delta x/t$), where $\Delta x$ is the distance from the boundaries,
$C_s$ is the sound speed evaluated using the source temperature, and $n_{ab}$ is the peak density
reached in the source region.  The simulation uses two species, carbon ($Z=6$) and
electrons, with heavy electrons ($Z m_e/m_C = 1/100$), compared to the physical mass ratio, for computational reasons.
The domain is $[-L,L]$ along $x$ and $[0, 2L/3]$ along the transverse direction,
which is included to allow multiple wavelengths of the Weibel instability to
grow.   We approximately match the ion-scale dimensionless parameters $L/d_{i,ab} \sim$~180 (expt) vs 130 (sim)
and $t_{laser} C_{s0} / L \sim$~0.17 (expt) vs. 0.21 (sim).  ($d_{i,ab}$ is the ion-skin depth calculated using the ablation density.)
Inter-particle collisions are modeled 
using a Monte-Carlo binary collision operator, with the collisionality 
chosen so that $\nu_{ei} / \gamma_{weib} \sim 10$ during instability growth, as estimated in the experiment.
The simulations used approximately \sciexp{5.7}{9}~computational particles.}
\label{FigWeibPIC}
\end{figure*}

The observed wavelengths are also in reasonable agreement with
Weibel-instability theory for DRACO-predicted plasma densities at the midplane.
Following Fig~\ref{FigWeibelTscan}(b), $\lambda_{max}/d_{iCH} \approx1$ for the expected range of 
ion temperature, implying $d_{iCH} \approx 150~\mu$m.
To compare, DRACO simulations find densities near \sciexp{2}{25}~m$^{-3}$ 
at the midplane for $t=$~3.8~ns, corresponding to $d_{iCH} \approx 65$~$\mu$m.
This is in reasonable agreement, and indicates that the observed
filaments are certainly on the ion scale.
(In contrast, they are much too large to be explained by the beam-driven
electrostatic instability, also present in the counterstreaming flow geometry \cite{KatoPoP2010}, but
which has characteristic wavelengths on the much smaller electron scale $d_e = (m_e / \mu_0 n_e e^2)^{1/2}$.)
The remaining factor 2--3 disagreement in wavelength may be explained 
by a combination of factors: (i) the possibility that DRACO is over-predicting density at the midplane,
possibly related to how the true elliptical laser focus
is converted to a cylindrically symmetric profile required by DRACO, (ii)
limitations of the local dispersion relation which excludes nonlocal effects,
(iii) the possibility that non-linear filament merging has already begun \cite{FonsecaPoP2003, SpitkovskyApJL2008}.

Finally, the observations are in agreement with particle-in-cell simulations.
The ablation flow geometry of the experiment is generated by seeding plasma to small volumes at the left
and right boundaries of the computational domain, which is initially in vacuum.
Figure~\ref{FigWeibPIC} shows the evolution of the plasma density and 
the development of magnetic filaments at the midplane due to the Weibel instability as the two flows
interpenetrate.  (The setup of the simulations is discussed in detail in the figure caption.)
The Weibel-generated fields 
grow and saturate on comparable timescales to the experiment, measured in units of the
dynamic time $L/C_s~(\approx 11$~ns in experiment), where $L$ is the target half-separation and $C_s$ is the sound speed.
The growth rates and characteristic wavelength of the modes 
at the midplane are measured directly in the simulations and 
are in reasonable agreement with the same ion-driven linear Weibel theory \cite{DavidsonPoF1972}.
The simulations predict peak Weibel-generated fields of order 
20~T, using the DRACO-predicted ablation parameters, 
giving B field energy approximately 1\% of equipartition with the
flow energy (c.f. \cite{KatoApJL2008}).   This is in reasonable agreement with the
proton caustic formation by magnetic deflection \cite{KuglandRSI2012}, which 
requires $\nabla_\perp \int \mathbf{B} \times d\mathbf{\ell} \sim$~60~T for typical proton energies and the experimental
proton magnification factors.  Here the line integral is along the proton trajectories and the gradient is taken in the object plane.  
This value can be interpreted as an upper bound from caustic lensing ``by one filament.''
For cumulative lensing by multiple filaments, the required magnitude per filament is correspondingly lower.
Finally, an important point is that in this 2-d simulation, the 
filaments are the transverse $B$ component (out of the page).  Interestingly,
this component does not scatter the diagnostic proton beam, at least to
lowest order.  However, in reality the magnetic turbulence will consist
of a 3-d honeycomb of filaments \cite{FonsecaPoP2003} with additional magnetic field components (required
by $\nabla \cdot \mathbf{B} = 0$), which would produce an observable perturbation to the diagnostic beam.

This work has identified plasma stream filamentation due to a Weibel-type instability
between collisionless counterstreaming laser-produced plasma plumes,
and simultaneously modeled its growth and saturation with massively parallel particle-in-cell simulation.
This instability has been proposed to be a necessary ingredient in forming shocks
in otherwise collisionless unmagnetized plasmas.  
These results suggest that future experiments at greater system size and at greater energy may be able to 
observe and study fully-formed Weibel-mediated collisionless shocks and
study its consequences for particle energization.

The authors thank the OMEGA EP team for conducting the experiments.
The particle-in-cell simulations were conducted on the Jaguar and Titan 
supercomputers through the Innovative and Novel Computational Impact
on Theory and Experiment (INCITE) program.  This research
used resources of the Oak Ridge Leadership Computing Facility located in the
Oak Ridge National Laboratory, which is supported by the Office of Science of the Department
of Energy under Contract DE-AC05-00OR22725.
This work is supported by the Department of Energy under contract DE-SC0007168.
%
%
%
%
%

\end{document}